\documentclass[preprint]{aastex63}

\usepackage{xspace}

\newcommand{\nustar}{\textit{NuSTAR}\xspace}
\newcommand{\eu}{$^{155}$Eu\space}
\newcommand{\xspec}{{\sc xspec}\space}
\newcommand{\offset}{\textit{Offset}\space}
\newcommand{\slope}{\textit{Slope}\space}
\newcommand{\degree}{$^{\circ}$}
\newcommand{\gauss}{{\tt gauss}\space}

\revised{\today}
\shorttitle{NuSTAR Gain Evolution}
\shortauthors{Grefenstette et al.}
\graphicspath{{./}{figures/}}
\begin{document}

\title{Measuring the Evolution of the NuSTAR Detector Gains}

\correspondingauthor{Brian Grefenstette}
\email{bwgref@srl.caltech.edu}

\author[0000-0002-1984-2932]{Brian W. Grefenstette}
\affiliation{Space Radiation Lab \\
California Institute of Technology \\
1200 E California Blvd \\
Pasadena, CA 91125, USA}

\author{Murray Brightman}
\affil{Space Radiation Lab \\
California Institute of Technology \\
1200 E California Blvd \\
Pasadena, CA 91125, USA}

\author{Hannah P. Earnshaw}
\affil{Space Radiation Lab \\
California Institute of Technology \\
1200 E California Blvd \\
Pasadena, CA 91125, USA}

\author{Karl Forster}
\affil{Space Radiation Lab \\
California Institute of Technology \\
1200 E California Blvd \\
Pasadena, CA 91125, USA}

\author{Kristin K. Madsen}
\affil{Space Radiation Lab \\
California Institute of Technology \\
1200 E California Blvd \\
Pasadena, CA 91125, USA}
\affil{Goddard Space Flight Center, Greenbelt, MD 20771, USA}

\author{Hiromasa Miyasaka}
\affil{Space Radiation Lab \\
California Institute of Technology \\
1200 E California Blvd \\
Pasadena, CA 91125, USA}



\begin{abstract}

The memo describes the methods used to track the long-term gain variations in the \nustar detectors. It builds on the analysis presented in \cite{Madsen_2015} using the deployable calibration source to measure the gain drift in the \nustar CdZnTe detectors. This is intended to be a ``live" document that is periodically updated as new entries are required in the \nustar gain CALDB files. This document covers analysis up through early-2022 and the gain v010 CALDB file released in version 20220510.

\end{abstract}



\section{Introduction} \label{sec:intro}

The \textit{Nuclear Spectroscopic Telescope ARray} \citep[{\em NuSTAR},][]{Harrison_2013} is a NASA Astrophysics Small Explorer observatory that launched in June of 2012. \nustar is composed of two co-aligned hard X-ray telescopes focused onto two focal plane arrays (hereafter FPMA and FPMB). We have previously described the time-dependent gain calibration of the detectors as a function of time \citep{Madsen_2015}. Here we describe an updated approach to the time-dependent gain and our process for validating our time-dependent gain corrections to determine if any additional corrections as a function of time are required.

The conversion of the pulse heights (PHAs) collected by the
\nustar detectors is a multi-stage process that relies on several
underlying CALDB files (for a full description see
the~\href{https://heasarc.gsfc.nasa.gov/docs/nustar/analysis/nustar_swguide.pdf}{\nustar
data analysis software users guide}). Several corrections were made to
the ``charge-loss correction" (CLC) files after launch to account for differences in the gain for the FPMA detectors compared to pre-launch estimates using the in-flight \eu calibration source. The primary gamma-ray decay lines are at high energy (86.54 and 105.4 keV), while there are several blended X-ray lines at low energy (including strong lines at 6.06 and 6.71 keV). In \cite{Madsen_2015} we used two epochs of the calibration source just after launch in 2012 and in January 2015 to determine the time-dependent change in the energy scale using the standard \xspec~formalism, where the transfer function to the correct \emph{PI}{~values is:}

\begin{equation}
PI^{\prime} = PI / Slope - Offset
\end{equation}

\begin{figure}
    \centering
    \includegraphics[width=0.7\textwidth]{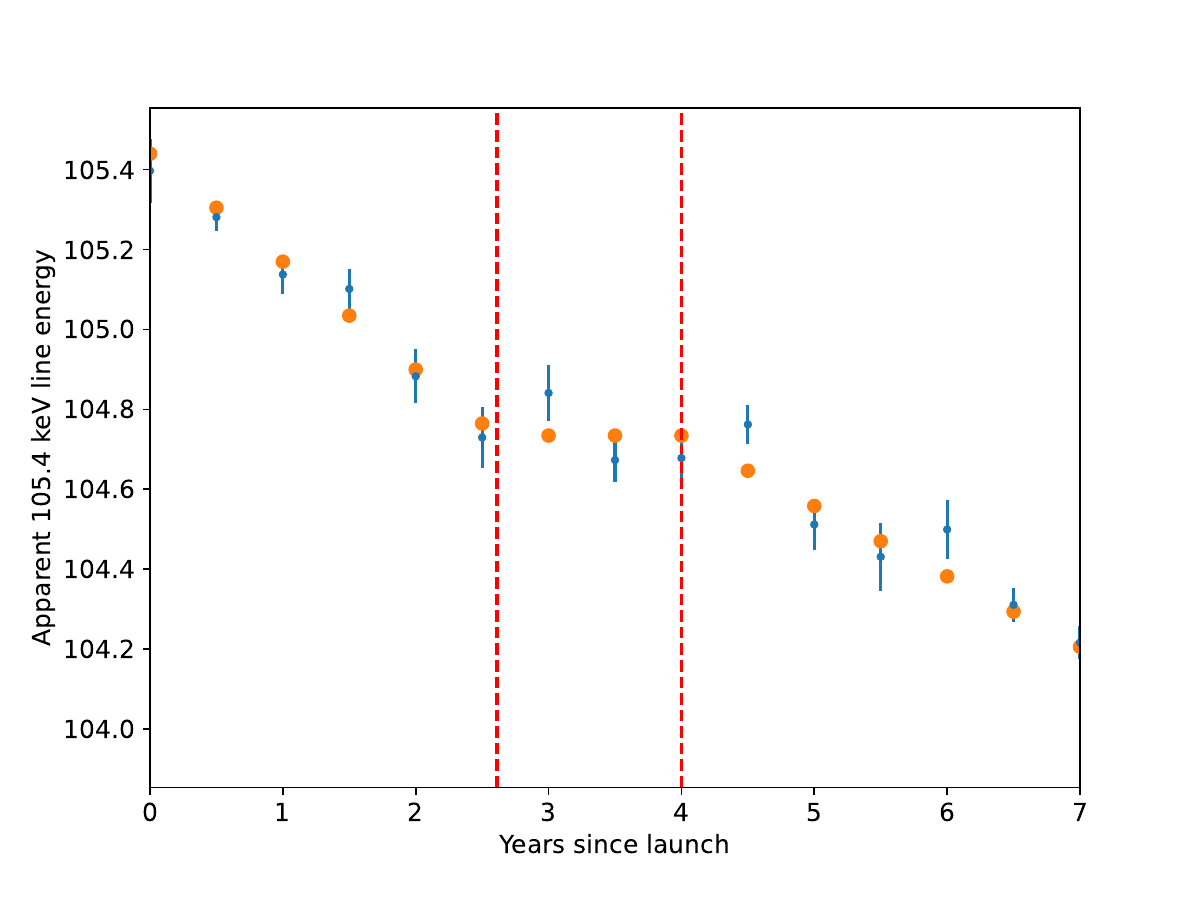}
    \caption{Long-term trend of the apparent energy of the 105.4 keV line for FPMA DET0 with all time-dependence in the calibration files removed. The blue data points and error bars represent the line centroid for each epoch, while the orange circles are the best-fit piecewise linear function. The vertical dashed lines show the epochs of the 2015 gain calibration and the end of the plateau in 2016. Only data through the 2019 analysis is shown here.}
    \label{fig:a0_trend}
\end{figure}

\ldots{}where in Table 7 of~\cite{Madsen_2015} we allowed both the \slope and
\offset values to vary as a function of time. Values were stored for the 2012
and 2015 epochs in the CALDB and the \texttt{NuSTARDAS} pipeline then
interpolates the corrections to a given epoch for an observation.

In subsequent analyses of observations of the Crab, we noted that allowing
the \offset parameter to vary with time was artificially
introducing low energy residuals in the Crab spectrum. In practice, this
implies that the charge-collection efficiency of the detectors may be
slowly varying with time, while the overall offset of the PI-to-keV transfer
function remains fixed.

We re-fit the 2012 and 2015 calibration source data with the \offset value
frozen to zero, thereby forcing all time-dependent variations into
the \slope parameter. This removed the artificial~low-energy
residuals in the Crab spectrum and was released in 2016. A full
history of the \nustar gain calibration files is given in Appendix \ref{ap:hist}.

The implementation of the gain CALDB files assumes that the variations
are linear with time with the measured changes in the gain corresponding
to a drop of 0.2\% per year from 2012-2015 and remaining flat after January
2015. \\

Further long-term monitoring of the the gain is accomplished without
risking the deployment of the source (which is mounted on a mechanical
arm that moves in and out of the field of view). Here we report on the two
methods used for monitoring the evolution of the \nustar gain: (1) monitoring the relative change in the~\emph{Slope} using
high-energy background lines in the \nustar background and (2) monitoring
any relative change in the \emph{Offset} using the Kepler supernova remnant.

\section{Monitoring long-term gain variations}

The \nustar internal background has several high-energy lines that we
can use to determine the variation in the detector gain with time. However,
as the \nustar background is also fairly low, this also requires a
substantial amount of integration time to measure the central energy of
the background lines.

We utilize all data where \nustar is pointed at the Earth (known
as \texttt{OCCULTED} or ``Mode 02'' in \texttt{NuSTARDAS} parlance). While the \nustar optics formally cut off at 78.4 keV, the detectors are capable of extending to much higher energies.
We use the data from 100 -- 150 keV, which are relatively easy to interpret
as the contributions mainly come from the internal background continuum component
as well as lines from the calibration source at roughly 105, 122, and 144 keV.

\subsection{2019 Update}

We reprocessed all of the data in the \nustar archive using the v004 CLC files
and the v005 GAIN files (see  Appendix \ref{ap:hist}) so that we can remeasure
the time dependence of the gain.

Data for each detector are integrated over six-month intervals, at which point we clean the data by requiring that the telescope bore-sight is more than 3\degree~below the horizon
to ensure that strong sources are fully blocked by the Earth. Standard
\texttt{OCCULTED} data files include times when the bore-sight is 3\degree~\textit{above}
the horizon, which is a conservative limit to avoid any attenuation in the Earth's atmosphere.
In addition, we require that the telescope is in Earth shadow to avoid any impact for solar activity.

For each epoch we then model the broad continuum using the standard broken power-law model \citep{Wik_2014} assumed to originate from down-scattered photons in the instrument. For this
source we use a diagonal response matrix file (RMF) (rather than the \nustar CZT RMFs that are
appropriate for photons normally-incident to the top surface of the detectors).
Since the continuum model is dependent on the space weather conditions (which vary with time) we allow the power-law indices above and below the break energy (fixed to 124 keV for convenience) as well as the normalization of the broken power law to vary. 

The line components are fit using several \gauss components, some of which phenomenologically
fit to the data (for $E>150$ keV) and some of which are known to originate from the radioactive
\eu source. The ones primarily used for our analysis have line energies of 105.4 keV and 144.6 keV. There are also strong lines at lower energies, but these are contaminated by internal
activation lines (which are also time-variable) in the CZT detectors and are not
usable for this analysis.

After we obtain a good fit for each epoch we use the Goodman-Weare \citep{goodman_ensemble_2010} MCMC implementation in \xspec and {\tt corner.py} \citep{foreman-mackey_corner.py:_2016} to estimate the 90\% quantiles and adopt these as the confidence limits for the 105.4  and 146.0 keV lines.

The trends for all eight detectors through mid-2019 are very similar. An example of the long-term gain trends is shown in Fig \ref{fig:a0_trend}. The initial drop of 0.2\% per year is confirmed as well as the flattening of the gain from 2015 to mid-2016. However, all eight detectors also show evidence of a further drop in the measured line energy from 2017 to 2019. We model this as a piece-wise linear fit, allowing the date of the two pivot points and the slopes before and after each pivot point to vary. Between the pivots the gain is assumed to be flat. The results are given in Table \ref{tab:bgdfit}. These changes in the gain were incorporated into the \nustar CALDB GAIN files as v008 in December, 2019.

The variations in the gain that we see are in family with the 0.2\% per year decrease during the first three years of the mission. DET0 (where most point source targets are located) has shown a drop of only $\sim$0.5\% and $\sim$0.3\% since 2015 on FPMA and FPMB, respectively. At 6.4 keV, this represents a drop of $\sim$30 eV, or less than one \nustar energy channel, but which may be detectable in extremely bright sources with strong Fe lines. The 20191202 CALDB release corrects even this small drop.

\subsection{2021 Update}

\begin{figure}
    \centering
    \includegraphics[width=0.7\textwidth]{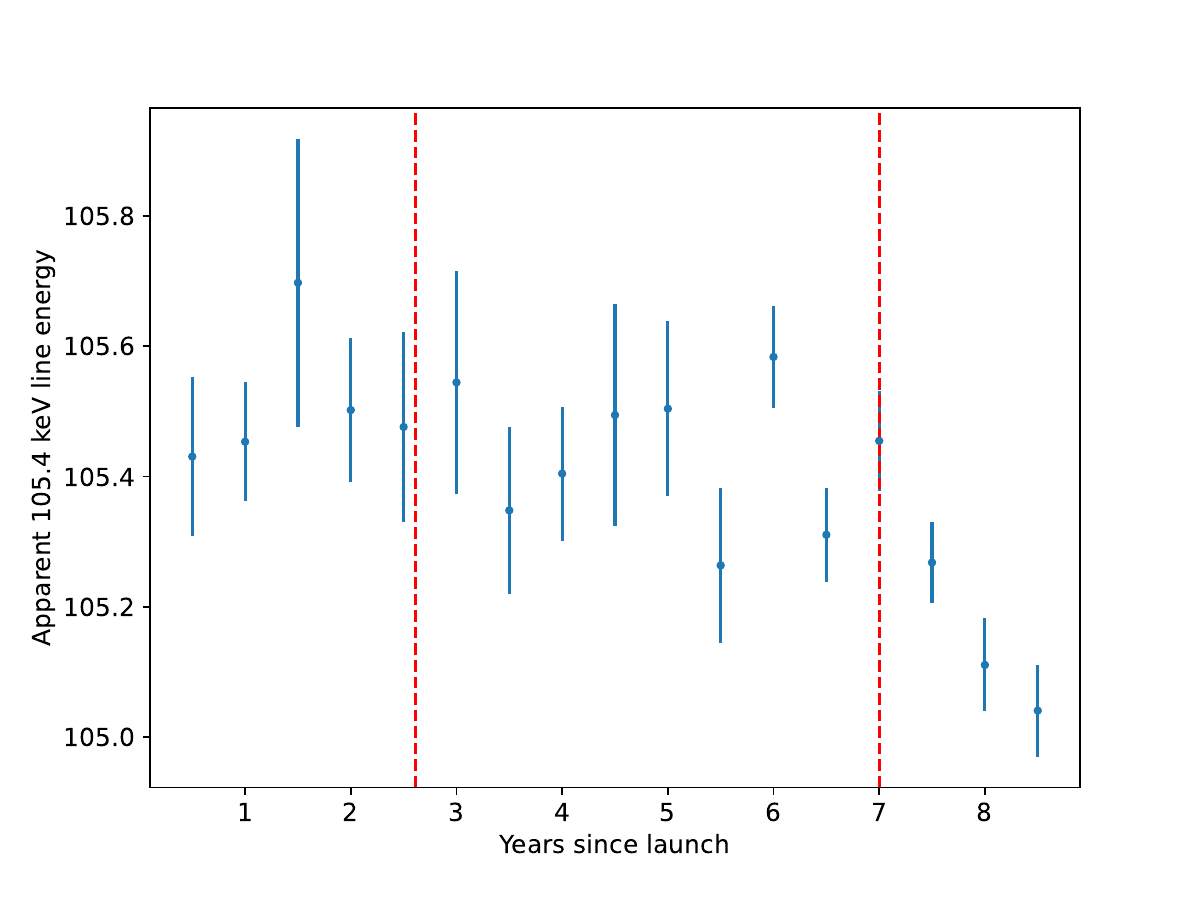}
    \caption{Long-term trend of the apparent energy of the 105.4 keV line using the v008 CALDB gain file for FPMA DET0. The vertical dashed lines show the 2015 and 2019 CALDB entries. All 8 detectors showed similar trends.}
    \label{fig:a0_2021_trend}
\end{figure}

\begin{figure}
    \centering
    \includegraphics[width=\textwidth]{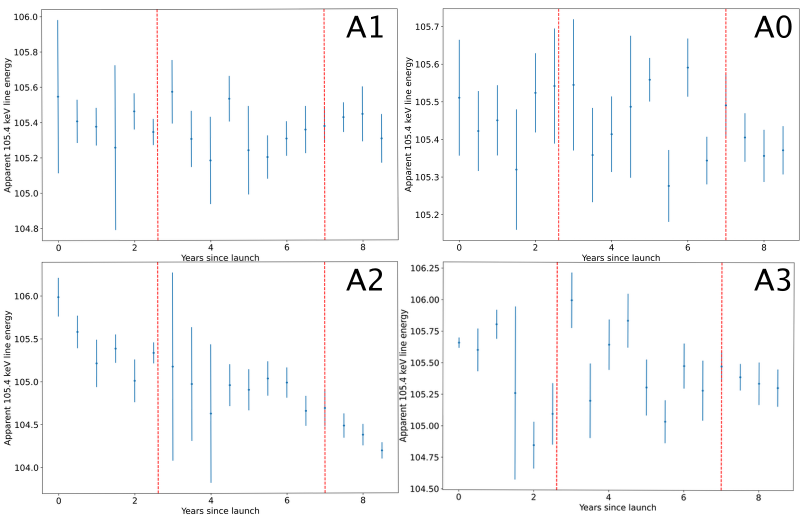}
    \caption{Apparent line centroids using CALDB v009 through July 2021 for the FPMA detectors.}
    \label{fig:fpma_summary}
\end{figure}

\begin{figure}
    \centering
    \includegraphics[width=\textwidth]{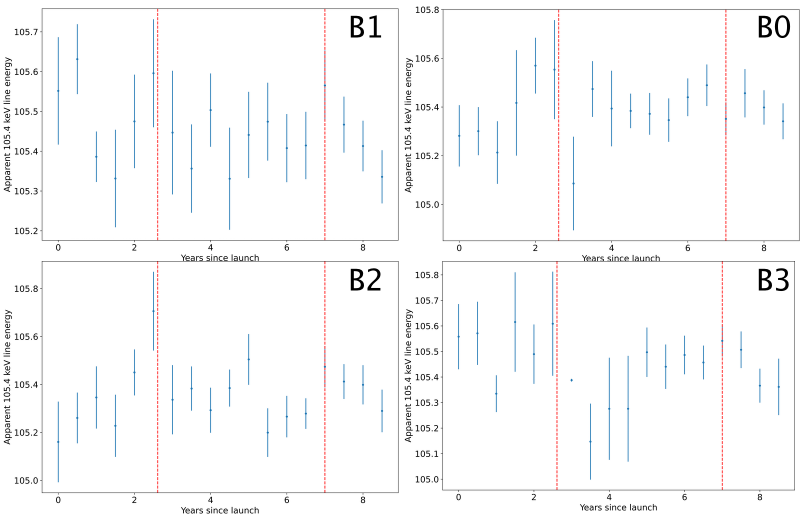}
    \caption{Apparent line centroids using CALDB v009 through July 2021 for the FPMB detectors.}
    \label{fig:fpmb_summary}
\end{figure}

We continued monitoring the apparent energy of the lines. By mid-2021 we noted a further drop in the line centroid location (Fig \ref{fig:a0_2021_trend}). All 8 detectors showed similar drops, so we applied a further gain correction in July of 2021 (Tab \ref{tab:bgdfit}, Figures \ref{fig:fpma_summary} and \ref{fig:fpmb_summary})

\begin{table*}
\begin{center}
\begin{tabular}{lrrrr}
\hline
Module & Det0 & Det1 & Det2 & Det3 \\
\hline
\multicolumn{5}{c}{Jan 2015 (v007) } \\
\hline
FPMA & 0.9933  & 0.9919  & 0.9993   & 0.9973  \\
FPMB & 0.9955  & 0.9929  & 0.9928   & 0.9960  \\
\hline
\multicolumn{5}{c}{Jan 2015-Sep 2016 plateau (v008)} \\
\hline
FPMB & 1.0* & 1.0*  & 1.0*  & 1.0*  \\
FPMA & 1.0* & 1.0* & 1.0*  & 1.0*  \\
\hline
\multicolumn{5}{c}{June 2019 (v008)} \\
\hline
FPMA & 0.995  & 0.994  & 0.995   & 0.993  \\
FPMB & 0.997  & 0.995  & 0.996   & 0.997  \\
\hline
\multicolumn{5}{c}{July 2021 (v009)} \\
\hline
FPMA & 0.996  & 0.994  & 1.0*   & 0.994  \\
FPMB & 0.996  & 0.996  & 0.994   & 0.998  \\
\hline
\multicolumn{5}{c}{December 2022 (v011)} \\
\hline
FPMA & 0.9967  & 1.0  & 1.0   & 1.0  \\
\hline
\end{tabular}
\caption{The epochs, CALDB version numbers, and the applied changes over time at each epoch starting with the v007 CALDB file. \\ **held fixed
\label{tab:bgdfit}
}
\end{center}
\end{table*}

\subsection{2022 Update}

\begin{figure}
    \centering
    \includegraphics[width=\textwidth]{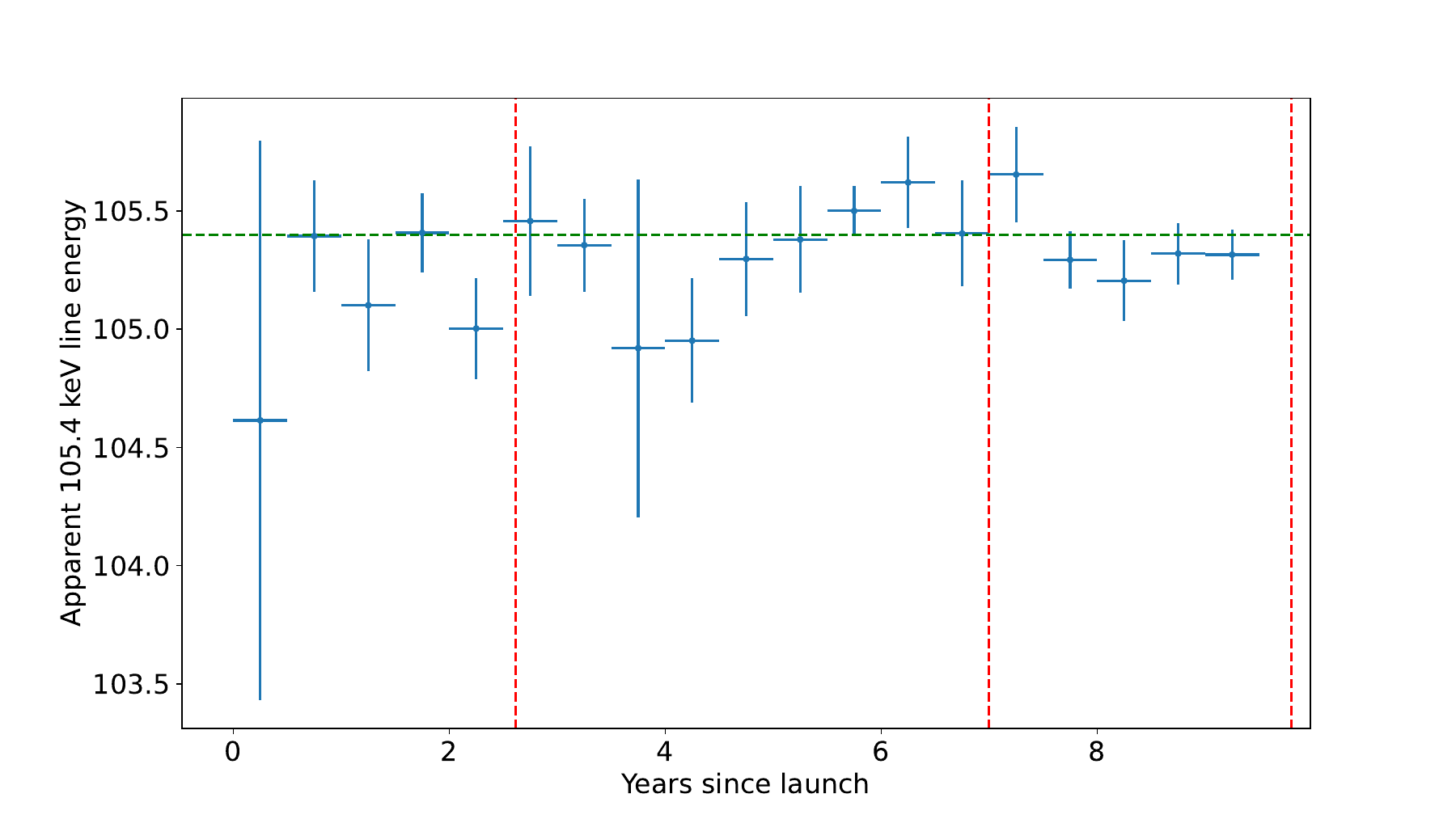}
    \caption{Apparent line centroids using CALDB v010 through May 2021 for the FPMA DET2 after the correction has been applied.}
    \label{fig:fpma2_update}
\end{figure}

We tracked the apparent energy of the line centroids through $\approx$ May of 2022 and found no need for an additional gain calibration point. We did, however, update the FPMA DET2 gain. This detector shows evidence for a linear decay in the gain from roughly 2015 to 2022, with the 105.4 keV line appearing at 98\% of its nominal flux by 2022. We added an entry in the gain calibration file for all pixels on this detector that accounts for this trend. Fig \ref{fig:fpma2_update} shows the line energies after the correction has been applied. This update was released in the nuAgain20100101v010.fits file associated with CALDB 20220504.

\subsection{Early 2024 Update}

We tracked the apparent energy of the line centroids through $\approx$ Jan of 2024 and found that only the FPMA DET0 required a new gain point in late 2022 with SLOPE change of 0.9967 (e.g., a drop of roughly 0.3\%). We added an entry in the gain calibration file for all pixels on this detector that accounts for this trend. This update was released in the nuAgain20100101v011.fits file associated with CALDB 20240226.

\subsection{Late 2024 Update}

We tracked the apparent energy of the line centroids through $\approx$ November of 2024 and found that none of the detectors required an additional data point.

\section{Offset Monitoring}

In the above section, we have explicitly assumed that the gain change is only in the \slope parameter. We can also use astrophysical sources to monitor the low-energy response of \nustar. Unfortunately, there are few astrophysical sources with reliable, clean, narrow line emission in the \nustar band-pass. Instead, we use the Kepler supernova remnant (SNR).

The remnant itself is resolved and the Fe-emiting regions have a range of Doppler velocities, making an absolute measure of the line centroids impractical for \nustar. The Fe is not uniformly distributed, so any changes in the effective exposure (e.g., due to detector gaps) can result in a shift in the measured spectrum. However, by observing the remnant at the same orientation and in the same location on the focal plane we can perform a simple search for \textit{relative} changes in the instrument response. 

\begin{figure}
    \centering
    \includegraphics[width=\textwidth]{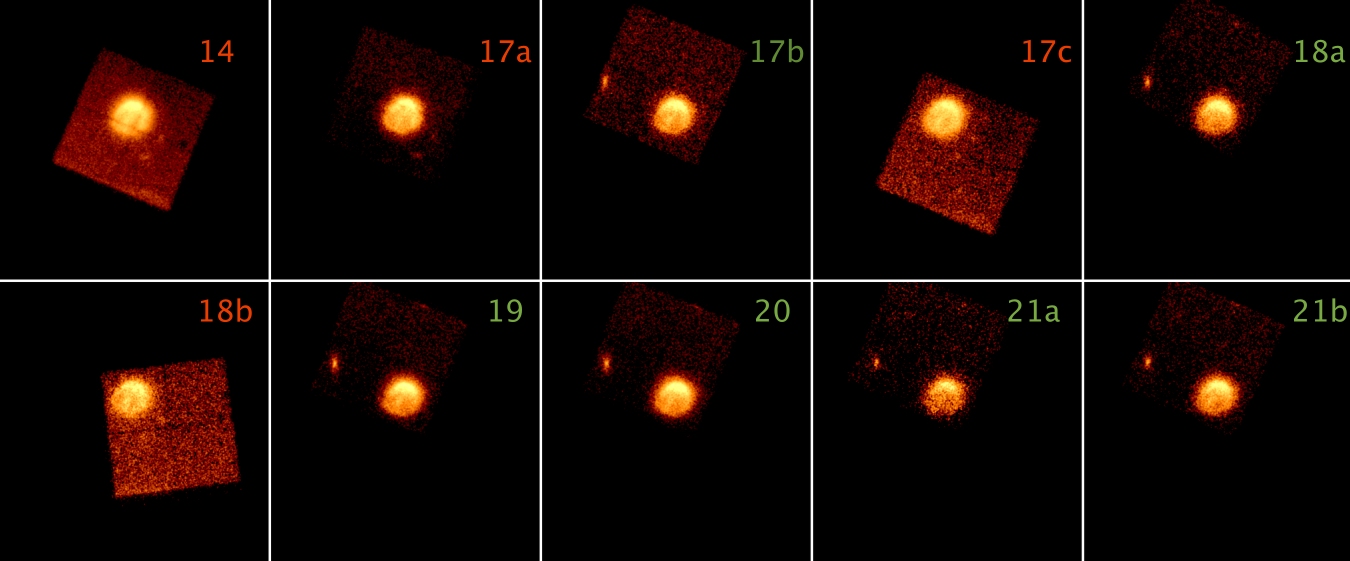}
    \caption{3--20 keV images from FPMA of Kepler for the observations listed in Table \ref{tab:kepler-obs}. Observations used in the model fits are shown in green. Epochs 22 and later were obtained at approximately the same roll angle as Epoch 21b and are omitted from this figure.}
    \label{fig:kepler_tiles}
\end{figure}

\begin{table*}[ht]
  \begin{center}
    \caption{Kepler Calibration Observations  \label{tab:kepler-obs}
}
    \begin{tabular}{llcccc} \hline \hline              
    Epoch & OBSID & Date & Exposure & PA & Used  \\ \hline
    14 & 40001020002 & 2014-10-11 & 246 ks & 336\degree & No \\
    17a & 90201021002 & 2017-02-07 & 126 ks & 158\degree & No \\
    17b & 90201021004	& 2017-04-22 &	51 ks & 156\degree & Yes \\
    17c & 90201021006 & 2017-10-08 & 54 ks & 336\degree & No \\
    18a & 90201021008 & 2018-06-04 & 42 ks & 147\degree & Yes \\
    18b & 90201021010 & 2018-06-17 & 47 ks & 6.2\degree & No \\
    19 & 10501005002 & 2019-03-17 & 100 ks & 158\degree & Yes \\
    20 & 10601005002 & 2020-03-18 & 100 ks & 157\degree & Yes \\
    21a & 10701407002 & 2021-05-04 & 18 ks & 155\degree & No \\
    21b & 10701407004 & 2021-05-10 & 50 ks & 155\degree & Yes \\
    22 & 10801407002 & 2022-04-27 & 100 ks & 154\degree & Yes \\
    23 & 10901407002 & 2023-05-04 & 112 ks & 155\degree & Yes \\
    24 & 11001407002 & 2024-04-17 & 95 ks & 155\degree & Yes \\
    
    \hline
    \end{tabular}
    \end{center}
\end{table*}

Table \ref{tab:kepler-obs} gives the full list of observations. While all of these data are useful for scientific analyses, for calibration purposes we only select sequence IDs when the remnant is fully clear of any detector gaps. After 2018, the calibration observations were designed to have the same position angle (PA) to ensure that no artificial differences are present in the remnant. Figure \ref{fig:kepler_tiles} show the observations listed in Table \ref{tab:kepler-obs}, where we are only using the observations where the entire remnant is located on DET0 on both FPMs (with the epochs annotated in green).

The remnant is compact (only 3-arcmin across), allowing us to easily integrate over the entire remnant. We concentrate on the low-energy X-ray band (2.5 -- 10 keV) where the spectrum can be described by a phenomenological model consistent of an absorbed, hot ($kT \sim$ 4 keV) \texttt{nlapec} along with seven \texttt{gauss} components. We note that the actual physics in the remnant is complex, so that the relative strength of the lines should not be physically interpreted from this simple model. We also note that over this energy range, the hot continuum component is essentially indistinguishable from a power-law. The atomic transitions are consistent with Si, S, Ar, Fe, and Ni. Except for the Fe-line complex, all of the lines are assumed to be narrow ($\sigma$ fixed to 0.1 keV). In the Fe line region the line is slightly broadened due to Doppler broadening of the Fe-rich ejecta in the remnant.

\subsection{Fitting procedure}

We first reprocess all of the Kepler data using the v009 CALDB files described above. This should remove any variations in the energy scale associated with the long-term \textit{Slope} drifts. All of the spectra are first rebinned using the \textsc{ftgrouppha} FTOOL and the optimally binning scheme from \cite{kaastra_2016}.

\begin{figure}
    \centering
    \includegraphics[width=0.49\textwidth]{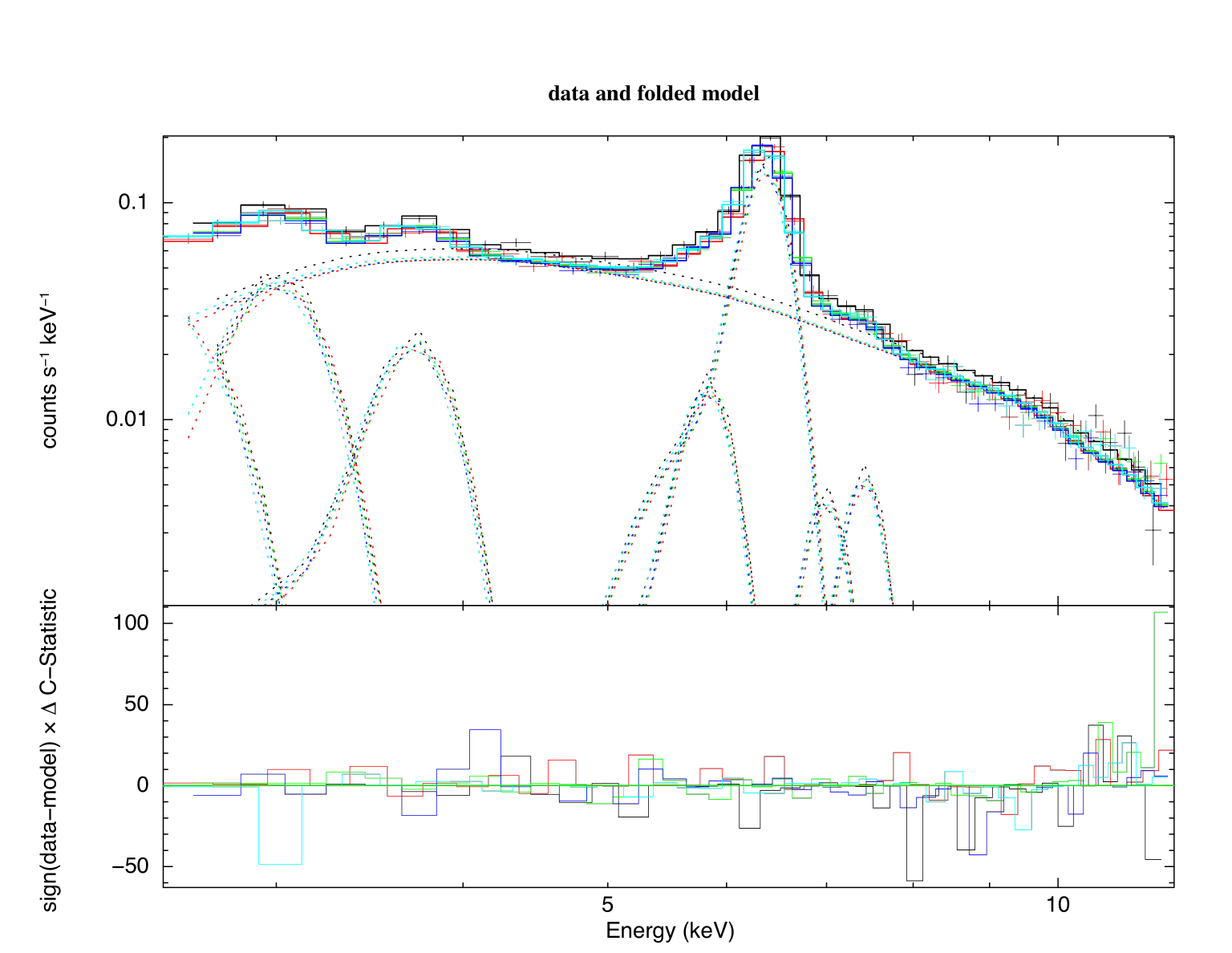} \includegraphics[width=0.49\textwidth]{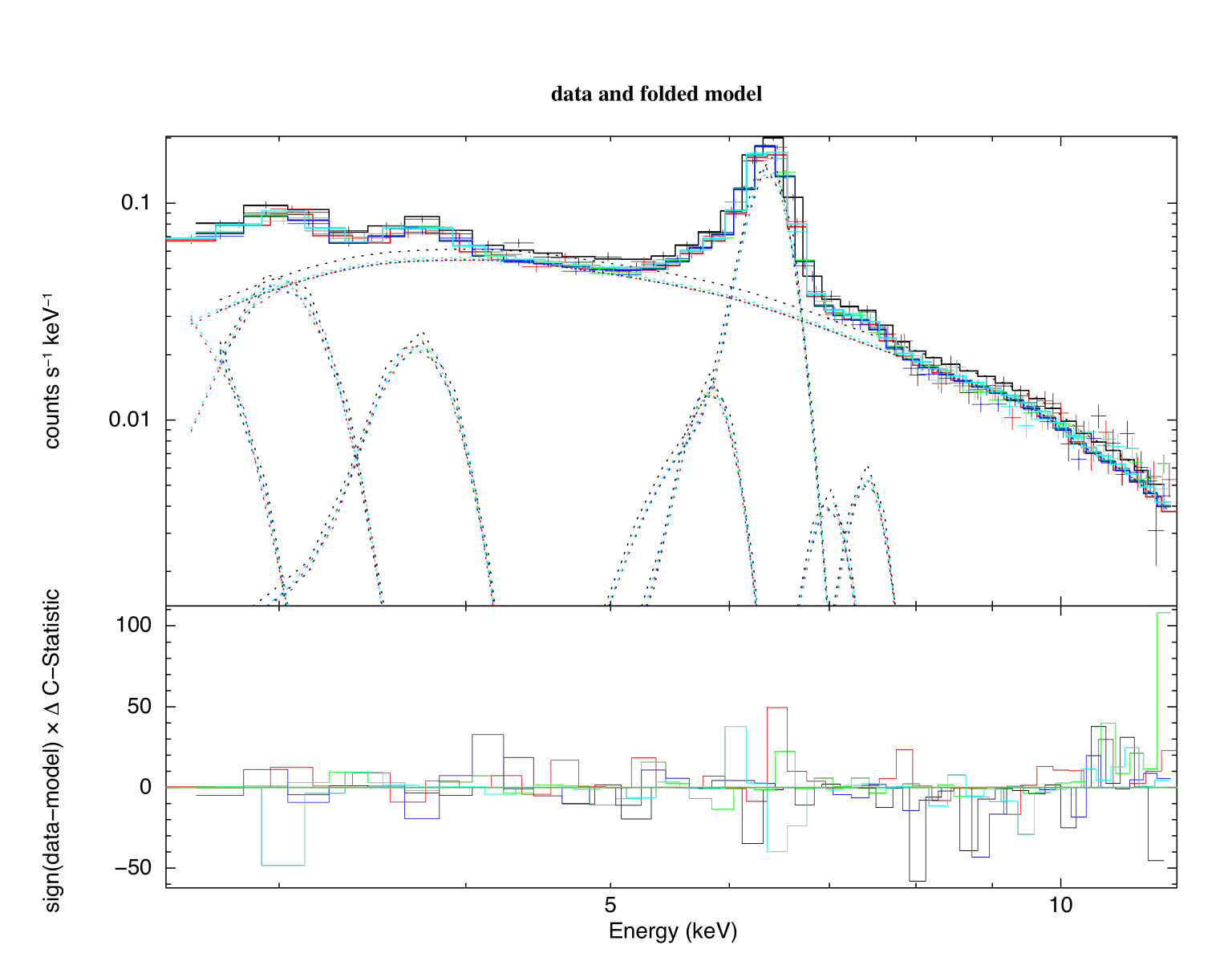}
    \caption{Spectrum of the Kepler SNR used to fit the gain. Each epoch is between 50-ks and 100-ks of exposure and has been ``optimally" binned (see text). The black, red, green, blue, and cyan data are for FPMB during epoch 17b, 18a, 19, 20, and 21b, respectively. The left panel shows the data after the gain fit has been applied, while the right figure shows after the gain fit has been applied. Epoch 22 has been included in the fits, but is not shown here.}
    \label{fig:kepler_gain}
\end{figure}

We simultaneously fit the six ``good" epochs, ranging from 2017 through 2022 using the 2.7--12 keV energy band using our ``standard" model. We arbitrarily pick Epoch 19 as a point reference since it has more exposure than Epoch 17b. We used the ``gain fit" formalism in \textsc{XSPEC}, freeze the \textit{Slope} parameter, and allow the \textit{Offset} to values to vary for each epoch and fit for the continuum parameters, the line fluxes and line centroids, and (for the Fe complex) the line width (Figure \ref{fig:kepler_gain}).

Table \ref{tab:kepler-fit} gives the \texttt{rerror} with a 2.706 delta-chi error range for the \textit{Offset} values for each of the epochs and for each FPM. All of the epochs have a fit value less than one \nustar channel (0.04 eV) and are consistent with the \textit{Offset} value being consistent with 0.0 except for Epoch 21b. While the fits are statistically improved when applying the gain fit (\texttt{cstat} 217/193 d.o.f. after the gain fit vs 250/195 before), it's not clear how such a small shift in the response edges is producing this significant change in the fit statistic.

All the same, we do note a slight trend in the energy with time at the 20 eV level. The scatter in the \textit{Slope} measurements above are on the order of 0.1 keV at 105.4 keV. Applying such a residual gain error at 6 keV results in an expected scatter on the order of 10s of eV. So it's possible that the residual trends are the result of un-modeled temporal changes in the \textit{Slope} parameter. Overall, we do not have significant evidence that the low-energy energy scale is changing with time on scales larger than the quoted systematic error of $\approx$ 40 eV (one \nustar channel).

.

\begin{table}[ht]
  \begin{center}
    \caption{Kepler fit parameters  \label{tab:kepler-fit}}
    \begin{tabular}{lrr} \hline \hline              
    Epoch & FPMA Offset (keV) &  FPMB Offset (keV)  \\ \hline
    17b & 0.02 $\pm$ 0.01 & 0. $\pm$ 0.1 \\
    18a & 0.03 $\pm$ 0.01 & 0.01 $\pm$ 0.01\\
    19 & 0* & 0.0* \\
    20 & -0.02 $\pm$ 0.01 & -0.01 $\pm$ 0.1 \\
    21b & -0.02 $\pm$ 0.01 & -0.03 $\pm$ 0.01 \\
    22 & -0.03 $\pm$ 0.01 & -0.02 $\pm$ 0.01 \\
    23 & -0.035 $\pm$ 0.01 & -0.02 $\pm$ 0.01 \\
    24 & -0.034 $\pm$ 0.01 & -0.025 $\pm$ 0.01 \\
    
    \hline
    \end{tabular}
    \end{center}
\end{table}

\appendix

\section{History of \nustar Gain-Related CALDB Files:}
\label{ap:hist}

Below are the documentation of the various \nustar CALDB files released since
launch. There are two relevant files, the ``pixel gain" (GAIN) file and the 
``charge loss correction" (CLC) file. At various times one or both of these have
been updated, so it's useful to keep track of these files here and for reference
in this paper.

\begin{itemize}
  \item nu[A/B]clc20100101v001.fits \textbar{} nu[A/B]gain20100101v004.fits
  
Pre-launch values based on ground calibration data and data obtained during environmental
testing (for FPMB).

  \item nu[A/B]clc20100101v002.fits  \textbar{} nu[A/B]gain20100101v005.fits

Released in ~July 2012 with updated based on 2012 in-flight calibration source data. No time dependence on \slope or {\em Offset}. Contains large corrections to FPMA, smaller corrections to FPMB. 

\item nu[A/B]clc20100101v003.fits \textbar{} nu[A/B]gain20100101v005.fits

Released in April 2014 and contain updated pixel-by-pixel \slope and \offset corrections to improve performance. Generally small corrections for most detectors, though large changes for FPMA DET2, which was not updated in the v002 data.

\item nu[A/B]clc20100101v004.fits \textbar{} nu[A/B]gain20100101v006.fits \\
Released in March 2015 and the first data set based on the 2015 \eu source deployment. The CLC file was update based on revised analysis of 2012 data and time-dependent \slope and \offset values were stored in the gain file.

\item nu[A/B]clc20100101v004.fits \textbar {} nu[A/B]gain20100101v007.fits

Released in July 2016. Reverts to using the 005 version of the gain file with the 004 version of the CLC and performs a cross-correlation between the 2012 and 2015 calibration data sets to determine the change in the \slope parameter. The \offset parameter is not allowed to vary with time.

\item nu[A/B]clc20100101v004.fits \textbar {} nu[A/B]gain20100101v008.fits

Released in December 2019 and based on the methods described in this paper.

\end{itemize}

\section{Internal list of gain files}
\label{ap:internal}

List includes CALDB files since the 2015 release, and only the gain files.

\begin{itemize}
  \item vcit011
  
  Did not track the intermediate plateau between 2015 and 2019 (unreleased)
  
  \item vcit012
  
  First fix of the bug (unreleased)
  
  \item vcit013
  
  Allowed FPMA DET3 to have a single linear slope across the time range (unrelased).
  
  \item vcit014
  
  Fits for FPMA DET2 gain drift in epoch 1 (unreleased)
  
  \item vcit015
  
  FPMA DET2 gain offset at launch ($\rightarrow$ released as version 008 in CALDB 20191213)
  
  \item vcit016
  
  2021 gain updates ($\rightarrow$ released as version 009 in CALDB 20211020)
  
    \item vcit017
  
  {\bf NOTE: As of this release, the versioning for FPMA and FPMB will be different.}   Contains updates to the FPMA DET2 gain described above ($\rightarrow$ released as version 010 for FPMA only) in CALDB 20220510.

    \item vcit018
  
  {\bf NOTE: FPMA only}
  
  Contains updates to the FPMA DET0 gain described above ($\rightarrow$ released as version 011 for FPMA only) released in CALDB 20240229.
  
  \end{itemize}
\bibliography{main}{}
\bibliographystyle{aasjournal}



\end{document}